%% file: scan-as-a-block-note.tex
\documentclass[11pt,letterpaper]{article}
\usepackage[margin=1in]{geometry}
\usepackage[T1]{fontenc}
\usepackage{booktabs,array}
\usepackage{graphicx}
\usepackage{tikz}
\usetikzlibrary{arrows.meta}
\usepackage{url}
\usepackage[colorlinks=true,linkcolor=black,citecolor=black,urlcolor=blue]{hyperref}
\graphicspath{{figures/}}

\title{Patchbay: a block model for the sonification of ADS-B aircraft data}
\author{Arquimedes Canedo\\Independent Researcher\\Sector Labs\\\href{mailto:canedo@sector.mx}{\texttt{canedo@sector.mx}}}
\date{25 October 2026}

\begin{document}
\maketitle

\begin{abstract}
Patchbay is a design for turning the data that an ADS-B receiver collects into sound by wiring typed blocks together. ADS-B reports are the position and altitude messages that aircraft broadcast and that a small ground receiver can collect. We first describe what such a receiver reports: signals that differ in shape (continuous, stepped, circular, categorical), in resolution, in who produces them, and in how long they stay put, which runs from about ten seconds to the whole time an aircraft is in view. From this we derive a small model. Four kinds of connection carry an aircraft set, a value, an event or audio, and 27 blocks select aircraft, read fields, decide when a note starts, shape values and make sound. One of these blocks makes the receiver act as a radar: scanning by bearing it is a line that rotates about the receiver, and scanning by range it is a ring that grows from the centre or shrinks inward. Each aircraft sounds at the moment the scan reaches it, and the scan's period is set in bars of a beat clock. The model does not depend on a browser. Our reference implementation is a browser editor, contained in one HTML file, that replays two recorded captures and records both an audio take and a MIDI take, a file of the notes played rather than of the sound. We say what is not built and place the work beside the closest systems we found, among them a radar-sweep sequencer for tangible objects from 2006 and a range-ring sonification of ADS-B aircraft published in 2026. Sweeps that trigger sound, and typed connections, both exist in earlier systems. To our knowledge, none of the sources we reached carries the aircraft in view, with every field attached, as one kind of connection in an editor.
\end{abstract}

\section{Introduction}

Patchbay is a tool for making music from aircraft. It is the editor of Pitchaxis, our project for sonifying ADS-B data, and it takes its name from the panel of jacks and cables with which a studio routes signals between devices. It has two parts. The first is a design: what the tool takes an aircraft to be, what can flow between blocks, which blocks exist and what each takes and gives, and how a graph of blocks is evaluated against a beat clock. The second is a reference implementation, a browser editor contained in one HTML file, in which a user chooses one of two recorded captures of the air traffic seen from a receiver in central New Jersey and builds a patch, a set of blocks joined by cables, on a canvas. Nothing in the design needs a browser. Section~\ref{sec:impl} says what another implementation would have to provide.

The data are not one signal. Section~\ref{sec:signals} characterizes what a receiver reports, signal by signal, because the choice of blocks depends on what the signals are like: how they are shaped, how finely they are resolved, how long they stay put and who produces them. Section~\ref{sec:model} gives the model that follows from that: four kinds of connection, the rules for joining them and the reasons for those rules, and 27 blocks, with what each takes and gives.

The idea we most want to describe is the scan block (Section~\ref{sec:scan}). A real radar turns a beam, and each aircraft is seen at the moment the beam passes it. The ADS-B receiver is not a radar: it only listens, it sends nothing and it has no beam. The scan block makes it act as one in the sound. In Patchbay the beam is a block that the user places and wires, centred on the receiver, and the moment it crosses an aircraft decides when that aircraft sounds; neither the arrival of a data sample nor a fixed clock does. Scanning by bearing, the block is a line that turns once per period. Scanning by range, it is a ring that grows from the centre, or shrinks inward, over one period. The period is counted in bars of the patch's beat clock, so the scan keeps time with the music.

This note describes the tool as it is built. It is not a user study, and it does not claim that a scan helps anyone do a task. It does not claim to be the first to sweep a line or a ring across tracked objects, across aircraft or across an image of the sky, or to put a sweep inside a patching environment; Section~\ref{sec:related} says who did so earlier. The contributions are small and specific:
\begin{itemize}
\item a characterization of the signals an ADS-B receiver reports, by shape, resolution, time scale, presence and origin, with a statement of which of them the tool carries;
\item a model of four kinds of connection and 27 blocks, with the reasons for its rules;
\item the scan, by bearing or by range, as a block that works on an aircraft set;
\item a worked mapping, the Flight levels example, with a check of the premise it rests on, and a browser implementation that replays recorded receiver data and records an audio take and a MIDI take.
\end{itemize}

\section{What the receiver senses}\label{sec:signals}

An ADS-B receiver is a radio that listens for the messages aircraft broadcast about themselves, among them position and altitude. The one we use runs readsb, a decoder that turns those messages into a list of tracked aircraft and writes the list to a file, \texttt{aircraft.json}, about every five seconds~\cite{wiedehopf2026readsb}. We call each written list a frame, and each aircraft's entry in it a row. A row is readsb's tracked state for one aircraft, not a raw radio message.

To decide which blocks to offer, we measured what such a receiver reports. We call the measurement the census: a recording of 356 frames over 1,797~s, made with the same receiver as the two captures the tool replays (Section~\ref{sec:impl}) but separate from them, and committed to version control on 29 September 2026. It lists 363 distinct aircraft, and its unit of counting is the aircraft-frame: one aircraft listed in one frame.

readsb goes on listing an aircraft for some time after it was last heard, and the values listed for such an aircraft are frozen. The census therefore counts an aircraft in a frame only if it was heard in the last 5~s. On that rule 73 percent of the listed aircraft-frames count, which is 92.9 of the 127.5 aircraft listed in a frame on average.

\paragraph{Origin.} The signals come from four places. The receiver makes some of them: signal strength and message rate are measured, and range and bearing are computed from the aircraft's position and the antenna's, so a different antenna reports different values for the same aircraft. The aircraft broadcasts others from its own GPS and instruments: altitude, climb rate, ground speed and track. The pilot or a controller sets a third group: the values dialled into the autopilot and the altimeter, the autopilot modes and the transponder code, an identifying code assigned by control. A fourth group is not broadcast at all but looked up in a database by the aircraft's address, the unique identifier it broadcasts: the aircraft type and the operator.

\paragraph{Shape.} Six shapes occur. A continuous signal, such as signal strength, can take any value within its range. A stepped signal, such as altitude, sits on a fixed grid, and the step is its resolution. An angle, such as bearing, wraps, so that 359 degrees is next to 1. A category, such as the aircraft type, is a label with no order and nothing between labels. A switch is on or off, or is a single event, and a combination, such as the autopilot modes, is several switches read together. Table~\ref{tab:signals} gives, for each signal, the shape, the range and resolution, and the share of aircraft-frames that carry it.

\begin{table}[t]
\centering\small
\input{figures/signals-table}
\caption{The signals of the census recording, by origin. Range is the span of the middle 98 percent of readings. Bits measure how evenly the values occur; one bit is an even two-way split. Present is the share of the 33,063 aircraft-frames in which the aircraft was heard that carry the signal; for a combination it is the share that carry all of its members.}
\label{tab:signals}
\end{table}

Four observations from the census shaped the design.

\begin{figure}[t]
\centering
\includegraphics[width=\linewidth]{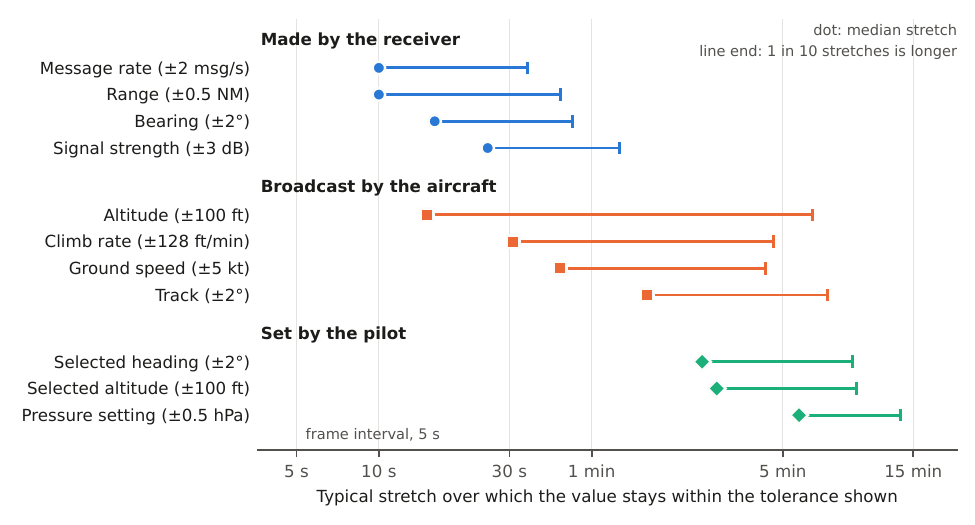}
\caption{How long each numeric signal stays put in the census recording. A dot marks the median stretch and the end of the line marks the length that one stretch in ten exceeds. Lengths are weighted by duration: the value plotted is the length of the stretch that a randomly chosen moment falls in. Every median is longer than the 5~s frame interval.}
\label{fig:timescales}
\end{figure}

\paragraph{Time scales differ by a factor of thirty or more.} Figure~\ref{fig:timescales} shows how long each numeric signal stays within a tolerance of where it was. The tolerances are chosen per signal and printed beside it; each is the smallest change we judged worth a note, such as 3~dB of signal strength or 100~ft of altitude. A stretch is the time a signal stays within its tolerance; it ends when the value leaves the tolerance or a frame is missing. The typical stretch is 10~s for message rate and range, 16~s for bearing, 25~s for signal strength, between 15~s and 96~s for altitude, climb rate, ground speed and track, and between 152~s and 344~s for the three values the pilot has selected. The identity labels (size class, type and operator) did not change in any aircraft while it was in view. The spread within a signal is also wide: one altitude stretch in ten lasts 385~s or more.

Two consequences follow for a tool that makes sound. A note started by the arrival of data would repeat on the receiver's clock, every five seconds, whatever the aircraft did. A note started by a change would be rare for the slow signals. So the tool treats when a note starts as a separate question from what the data say, and gives it its own kind of connection and its own blocks (Section~\ref{sec:model}).

\paragraph{There are fewer independent signals than Table~\ref{tab:signals} lists.} In this sky the high aircraft are also fast and far from the antenna, so aircraft that rank high in altitude also tend to rank high in ground speed and range. For each aircraft we took the median, over its time in view, of each of 20 channels that carry numbers. For each pair of channels we then measured how strongly the two are associated across aircraft. The measure is the absolute rank correlation, or its circular counterpart when an angle is involved: how far the ordering of the aircraft by one channel matches their ordering by the other, in either direction. We then merged channels into groups step by step, each time joining the two groups whose members were on average most associated (average linkage), until the mean association inside a group fell below a cut. At a cut of 0.5 the 20 channels form 10 groups; at 0.3 they form 6 and at 0.7 they form 13. At 0.5, range, pressure altitude, GPS altitude, selected altitude, ground speed and pressure setting fall into one group. A patch that maps altitude to pitch and ground speed to timbre therefore draws twice on one source of variation. The association is measured across the aircraft in one recording and reflects that traffic, not a property of aircraft in general.

\paragraph{Labels differ in how much they tell.} Size class has 6 values (1.7 bits), aircraft type 77 (5.4 bits) and operator 95 (5.0 bits). We expect that telling 77 categories apart by ear is a hard task and that a split into a few classes is not. The editor therefore offers one categorical split, engine class (jet, turboprop or piston), which it looks up from the aircraft type, and offers neither type nor operator as a field. We have not tested with listeners that three classes can be told apart and 77 cannot; the choice is a design judgement.

\paragraph{The receiver sets limits.} Nothing faster than the 5~s frame is visible. Only 73 percent of the listed aircraft are heard in a frame. Some signals exist only for some aircraft: the GPS climb rate is present in 31 percent of aircraft-frames and the selected heading in 60 percent, and indicated airspeed, Mach number and bank angle appear for about 1 percent of aircraft. The source of a contact, the way the receiver came to know of an aircraft (ADS-B, a Mode S reply with no position, or a position worked out from the arrival times of its signal at several receivers), changed 560 times across 351 aircraft as reception came and went.

\paragraph{What the tool carries.} The tool carries eight of these signals for every aircraft: altitude, climb rate, ground speed, track, range, bearing, signal strength and message rate. It also carries four labels: callsign, aircraft type, engine class and wake class, a class by weight. The values the pilot has selected, the transponder code, the source of the contact and the quality codes are in the census and not in the tool. Section~\ref{sec:limits} lists them among the things not built.

\section{The model}\label{sec:model}

\subsection{Four kinds of connection}

\begin{figure}[t]
\centering
\resizebox{\linewidth}{!}{\input{figures/fig-kinds}}
\caption{The four kinds of connection and the blocks that join them. An arrow names the blocks that take one kind and give another. A box names the blocks that take a kind and give the same kind. Sky, Number and Every have no input and start a kind; Output ends the audio. Closest pair and Level-off appear more than once because each gives more than one kind: Closest pair an aircraft set, values and events, Level-off an aircraft set and an event. Ground appears twice: it keeps the value kind, and it is the one block that takes an event and gives values.}
\label{fig:kinds}
\end{figure}
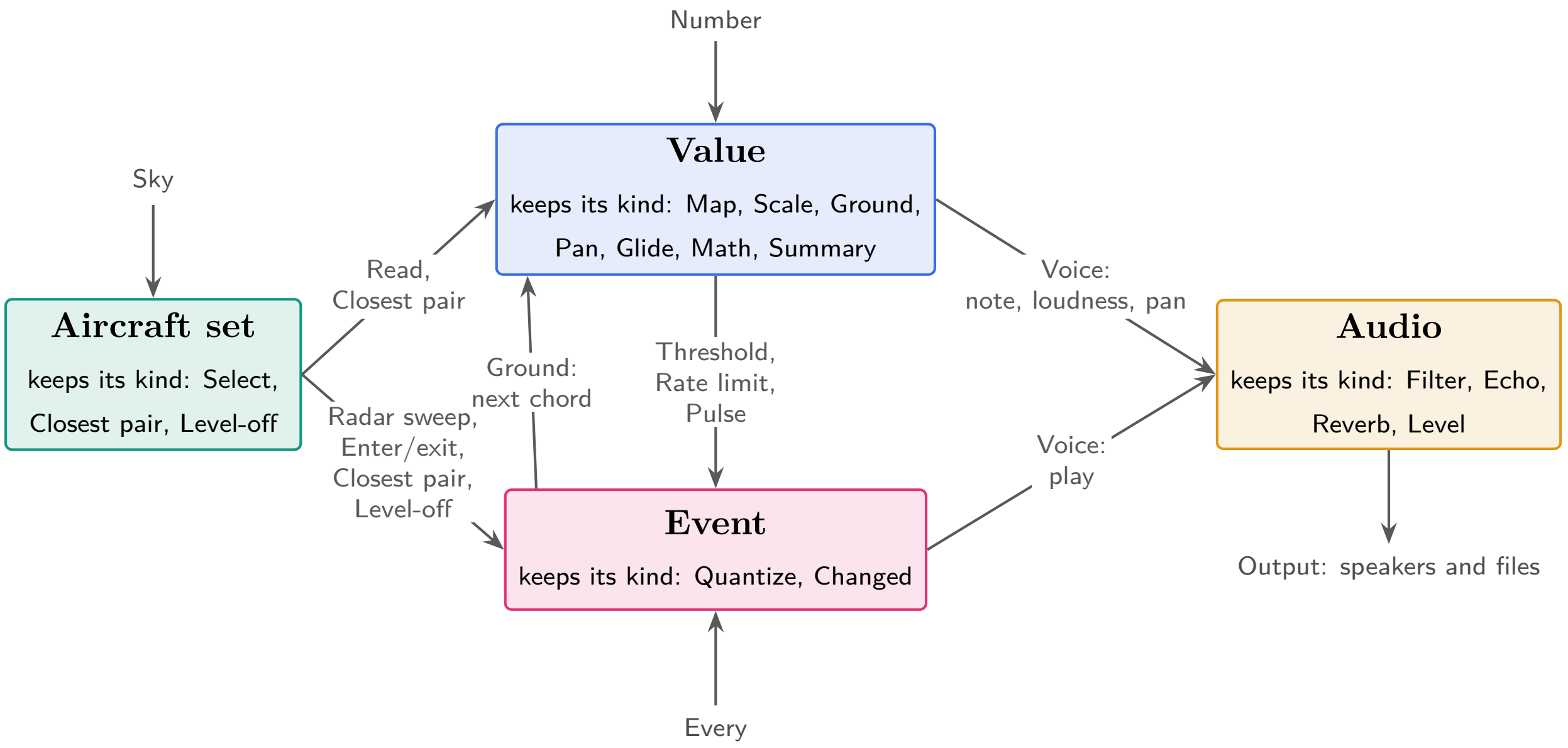

A patch has to answer four questions: which aircraft are in play, what number goes into each sound, when each sound starts, and what it sounds like. Patchbay gives each question its own kind of connection, and we call these four the kinds (Figure~\ref{fig:kinds}). An aircraft set carries the aircraft in view, each with every field attached, a field being one of the eight signals or four labels that the tool carries. A value carries one number for each aircraft, or one number for all. An event is a moment in time together with the aircraft that caused it. Audio is sound. A cable joins an output of one block to an input of another of the same kind.

\subsection{Why the rules are what they are}

The rules for joining blocks follow five decisions.

\begin{figure}[t]
\centering
\includegraphics[width=0.52\linewidth]{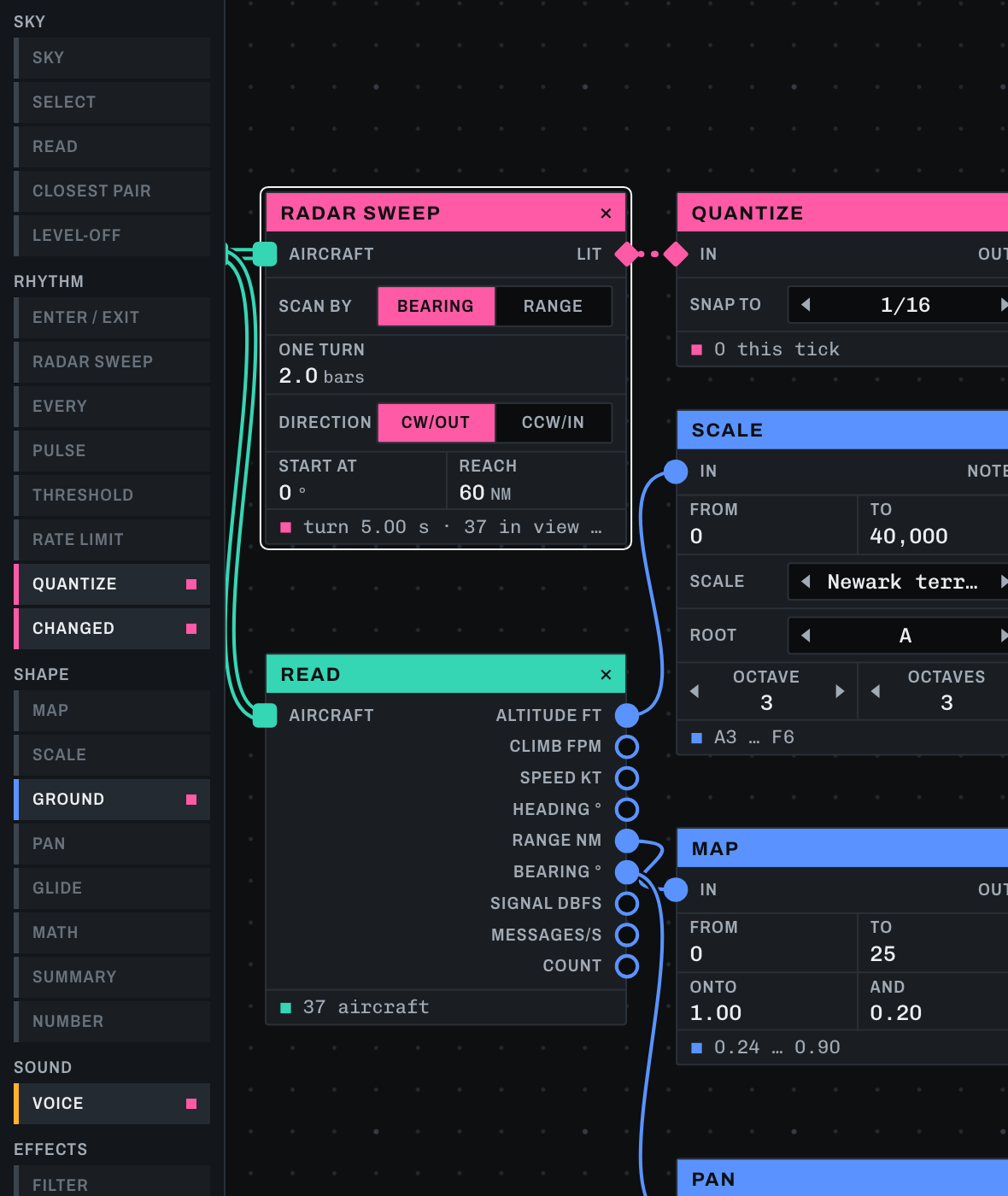}\\[2pt]{\small (a)}\\[8pt]
\includegraphics[width=0.7\linewidth]{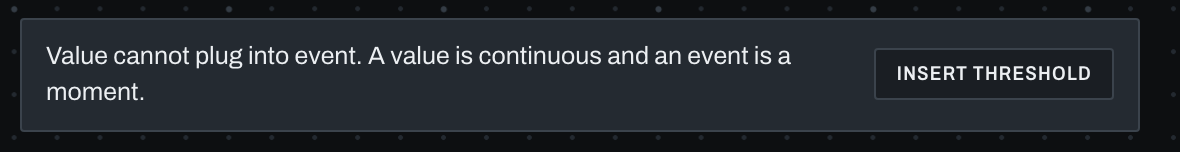}\\[2pt]{\small (b)}
\caption{The typing is shown before it is enforced and explained when it refuses. (a) With the Radar sweep selected and outlined, the library on the left marks the four blocks that can take its event output, Quantize, Changed, Ground and Voice, with a pink square. (b) Dragging a value cable to an event input is refused, with the reason and an offer to insert the block that bridges the two, Threshold.}
\label{fig:cue}
\end{figure}

\begin{enumerate}
\item \textbf{The aircraft set travels whole, and values are keyed by aircraft.} A patch that gives each aircraft its own voice must know, at the Voice, which note, loudness and stereo position belong to the same aircraft. If a cable carried bare numbers, the three would have no way to meet again. So Select keeps or drops whole aircraft, a field is never separated from its aircraft, and Read opens a set into one value per field, each a number for every aircraft, labelled with the aircraft. A value can also be one number for all, as in a count or a constant.

\item \textbf{When a note starts is a question of its own.} The census showed that the data change more slowly than the frame that delivers them (Section~\ref{sec:signals}). The moment a note starts is therefore not read from the data but comes from a block that gives an event: most are in the Rhythm group (Table~\ref{tab:blocks}), and Closest pair, Level-off and Ground give events too. A Voice in held mode needs no event: every aircraft in the set holds its own tone, and the tone follows the value.

\item \textbf{A join is checked when it is made, and a refusal explains itself.} The editor refuses a cable between kinds that differ and says why in words, for example that a value is continuous and an event is a moment. For three of the twelve pairs of differing kinds it also offers the block that would bridge them (Figure~\ref{fig:cue}b). It refuses a cable that would make a loop, because feedback is not supported. An input of the aircraft, value or event kind accepts one cable, so it never has two sources; a second cable replaces the first, with a notice. An audio input accepts any number of cables and mixes them. The typing is shown before it is enforced: selecting a block marks the blocks in the library that can be joined to it directly (Figure~\ref{fig:cue}a).

\item \textbf{Nothing fails silently.} Every block shows a live readout of what it is doing: how many of the arriving aircraft Select kept, the range of notes that Scale produces, how many events a second a block fires. A block with an unwired input says what is missing. Map counts and reports the values it clamps, and Voice reports the notes it dropped at its polyphony ceiling. A block that raises an error shows the error and gives no output, and the rest of the patch goes on.

\item \textbf{The typing stops at the kind.} A value cable carries a number, with no unit and no shape. The census shows what this leaves out: bearing and track are angles, altitude sits on a 25~ft grid and a label has no order. The blocks that need to know handle it themselves. Pan takes the sine of a bearing, so that 359 and 1 degrees sit together; the scan compares angles modulo 360; and the interpolation that places each aircraft between two frames of the capture turns a bearing the shortest way round. Glide and Summary do not: Glide moves a value in a straight line toward its target, so a bearing that goes from 359 to 1 slides the long way round, and the mean that Summary takes of 359 and 1 is 180. Both are known faults. We intend a type to supply limits, units and warnings without ever rejecting a cable or rewriting a value the user has set; none of that is built (Section~\ref{sec:limits}).
\end{enumerate}

\subsection{The blocks}

\begin{figure}[t]
\centering
\begin{minipage}[b]{0.31\linewidth}\centering\includegraphics[width=\linewidth]{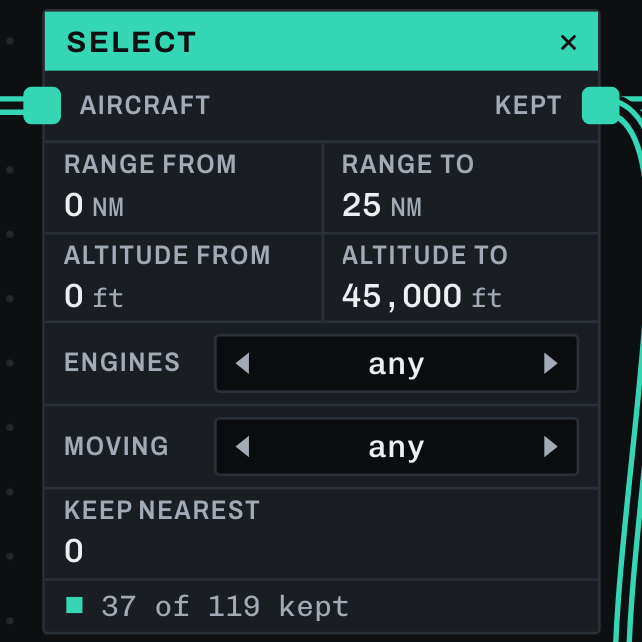}\end{minipage}\hfill
\begin{minipage}[b]{0.31\linewidth}\centering\includegraphics[width=\linewidth]{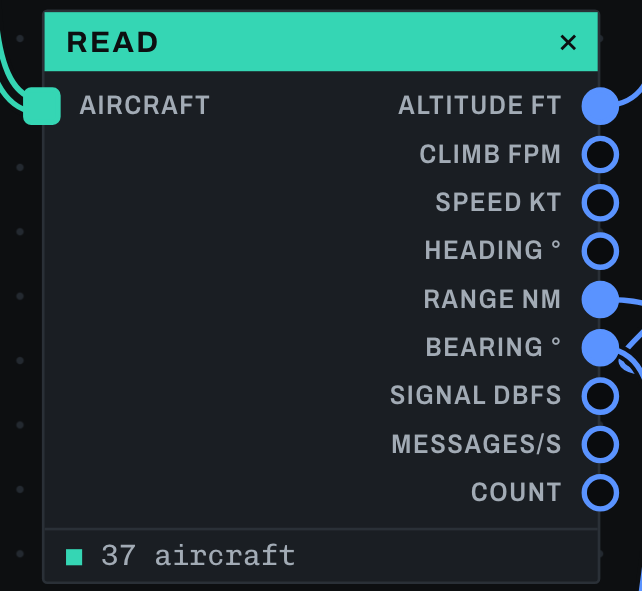}\end{minipage}\hfill
\begin{minipage}[b]{0.31\linewidth}\centering\includegraphics[width=\linewidth]{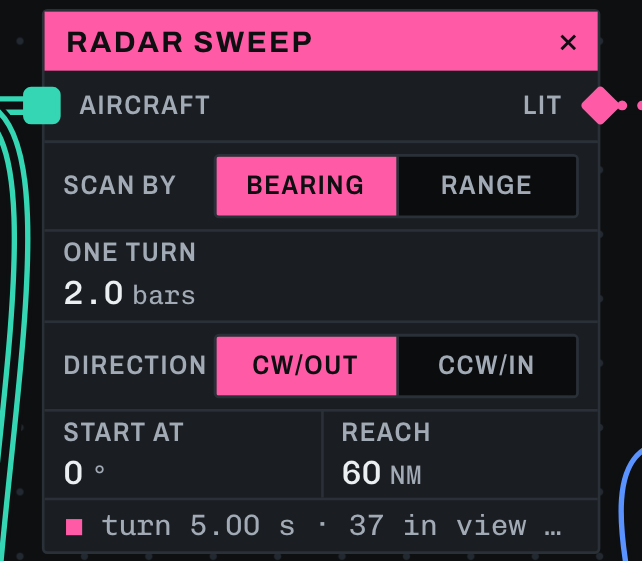}\end{minipage}
\caption{Select, Read and Radar sweep as the editor draws them, while a patch plays. Inputs are on the left and outputs on the right. The shape and colour of a jack give the kind of cable: a teal square for an aircraft set, a blue circle for a value and a pink diamond for an event. A filled jack is wired. Select keeps the aircraft that match its settings, Read gives one value per aircraft for each field, and Radar sweep turns an aircraft set into an event. The line at the foot of each block is its live readout.}
\label{fig:blocks}
\end{figure}

There are 27 blocks in five groups (Table~\ref{tab:blocks}). The groups follow the four questions: Sky answers which aircraft and which fields, Rhythm answers when, Shape answers what number, and Sound and Effects answer what it sounds like. The blocks that change kind are the adapters between the questions (Figure~\ref{fig:kinds}). From an aircraft set, Read gives values, and Radar sweep and Enter/exit give events. Closest pair gives both, for the one pair about to pass closest. Level-off gives an event and a smaller aircraft set, the aircraft that have just settled into level flight. From a value, Threshold, Rate limit and Pulse give events. Voice takes events and values and gives audio. Ground is the one block that takes an event and gives values: the event advances a chord progression to its next chord, and the values are notes (Section~\ref{sec:music}). Figure~\ref{fig:blocks} shows three of the blocks as the editor draws them.

\begin{table}[tp]
\centering\footnotesize
\newcolumntype{L}[1]{>{\raggedright\arraybackslash}p{#1}}
\begin{tabular}{@{}L{1.4cm}L{1.9cm}L{2.2cm}L{2.5cm}L{6.6cm}@{}}
\toprule
Group & Block & Takes & Gives & What it does\\
\midrule
Sky & Sky & none & aircraft set & replays the capture: the aircraft in view, every field attached; it can instead give the sky as it is a set number of seconds later\\
 & Select & aircraft set & aircraft set & keeps aircraft by range, altitude band, engine class, climb state, or the nearest $n$\\
 & Read & aircraft set & 8 values, a count & opens one field per aircraft: altitude, climb rate, ground speed, heading, range, bearing, signal strength, message rate\\
 & Closest pair & aircraft set & aircraft set, 3 values, 2 events & the one pair about to pass closest, found by predicting each pair's closest approach in a straight line; gives the pair, its distance, predicted miss and time to closest, and events for a predicted close pass and for the pass\\
 & Level-off & aircraft set & aircraft set, event & an aircraft that was climbing or descending settles level: an event, and the last few to do so as a set, the cast\\
\midrule
Rhythm & Enter/exit & aircraft set & 2 events & an aircraft joins the set; an aircraft has been gone longer than a hold time\\
 & Radar sweep & aircraft set & event & the scan front reaches an aircraft (Section~\ref{sec:scan})\\
 & Every & none & event & a fixed interval of beats; one event for all aircraft\\
 & Threshold & value & event & an aircraft's value crosses a level\\
 & Rate limit & value & event & an aircraft's value has changed by a set amount, no sooner than a floor and no later than a ceiling\\
 & Pulse & value & event & events at a pace a value sets, in beats; silent while there is no value\\
 & Quantize & event & event & moves each event to the next grid line\\
 & Changed & event, value & event & lets an event through only if its aircraft's value has moved since the last one let through, or a set time has passed\\
\midrule
Shape & Map & value & value & rescales one range onto another and counts the values it clamps\\
 & Scale & value & value & snaps a value to the notes of a scale\\
 & Ground & 2 values, event & 3 values, event & turns altitudes into the notes of a chord and moves to the next chord when an event arrives; gives a note per aircraft, a bass note, the chord's notes, and an event at each change\\
 & Pan & value & value & turns a bearing into a stereo position\\
 & Glide & value & value & smooths each aircraft's value over time\\
 & Math & 2 values & value & adds, subtracts, multiplies, or takes the minimum or maximum\\
 & Summary & value & value & one number for the whole set: mean, lowest, highest, spread or count\\
 & Number & none & value & a constant\\
\midrule
Sound & Voice & event, 3 values & audio & one voice per aircraft, in notes or held mode; takes play, note, loudness and pan\\
\midrule
Effects & Filter & audio, value & audio & low-pass, high-pass or band-pass; the cutoff may come from a value\\
 & Echo & audio & audio & an echo whose time is set in beats\\
 & Reverb & audio & audio & a room of a chosen size\\
 & Level & audio, value & audio & gain; a value may ride it\\
 & Output & audio & none & plays the mix and records it\\
\bottomrule
\end{tabular}
\caption{The 27 blocks, with what each takes and gives. An input accepts one cable, except an audio input, which mixes any number.}
\label{tab:blocks}
\end{table}

A typical patch runs from the Sky block to the Output block. Sky puts the aircraft in view on an aircraft-set cable. Select keeps the aircraft that match a range, an altitude band, an engine class, a climb state, or a limit on how many of the nearest to keep. Read opens an aircraft set into one value per aircraft for any of the eight signals the tool carries (its heading field is the track over the ground), or into a single count of the aircraft in the set. Map and Scale turn a value into another range or into the notes of a musical scale, and Pan turns a bearing into a stereo position. A Rhythm block decides when each aircraft sounds. Voice makes sound, and Output plays it.

\subsection{Evaluation}

A patch is evaluated in ticks, short steps of time. At each tick the sky at the current data time, the moment in the capture that the replay has reached, is computed once. Blocks are then evaluated in an order in which every block follows the blocks that feed it; such an order exists because loops are refused. Each block receives the outputs of the blocks wired to its inputs and returns its own. Every block is evaluated whether or not it reaches an Output. Each tick covers a window of beat time and of audio time, and a block that produces events stamps each one with a time inside the window. When the patch is not playing it is still evaluated, every 400~ms, in a preview in which the blocks that produce events produce none, so that readouts stay live without sound.

A patch has two clocks of its own, besides the audio clock, which is the real time of the sound output. The beat clock has a tempo, in beats per minute, and a number of beats per bar, and the defaults are 96 beats per minute and four beats per bar. The data clock, which says how far the replay has advanced through the capture, runs at the replay speed times the audio clock. Periods in a patch are set in beats or bars, and the replay speed is chosen separately.

\section{The scan block}\label{sec:scan}

The Radar sweep block decides when each aircraft in a set sounds. It takes an aircraft set and gives events, each one aircraft together with the time at which the scan reached it. Table~\ref{tab:scan} lists its settings.

The block is called Radar sweep because it makes the receiver act as a radar, which it is not: it listens and sends nothing, and it reports snapshots of the sky, not the returns of a rotating antenna. The block lays the scan over the data, and the scan is the one thing in a patch whose counterpart is not in the data. What the data do have is the geometry: range and bearing are measured from the antenna, so the sky is already in the polar coordinates a radar uses, the scan works in the receiver's own coordinates, and from an antenna elsewhere the same aircraft would have other ranges and bearings.

\begin{table}[ht]
\centering\small
\begin{tabular}{@{}llll@{}}
\toprule
Setting & Meaning & Default & Used by\\
\midrule
Scan by & bearing, a rotating line; or range, a ring & bearing & both\\
One turn & length of one turn, in bars (at least 0.25) & 2 bars & both\\
Direction & clockwise or outward; counter-clockwise or inward & clockwise, outward & both\\
Start at & starting bearing, in degrees & 0 & bearing\\
Reach & where the ring ends, in nautical miles (at least 5) & 60 & range\\
\bottomrule
\end{tabular}
\caption{Settings of the Radar sweep block.}
\label{tab:scan}
\end{table}

The length of a turn is set in bars of the patch's beat clock rather than in seconds. One turn lasts a whole number of bars, or a fraction down to a quarter bar, and its length in seconds follows from the tempo: with the defaults, two bars of four beats at 96 beats per minute is five seconds. At the default replay speed of four, one five-second turn then covers 20~s of the capture.

\begin{figure}[t]
\centering
\begin{minipage}[b]{0.43\linewidth}\centering
\includegraphics[width=\linewidth]{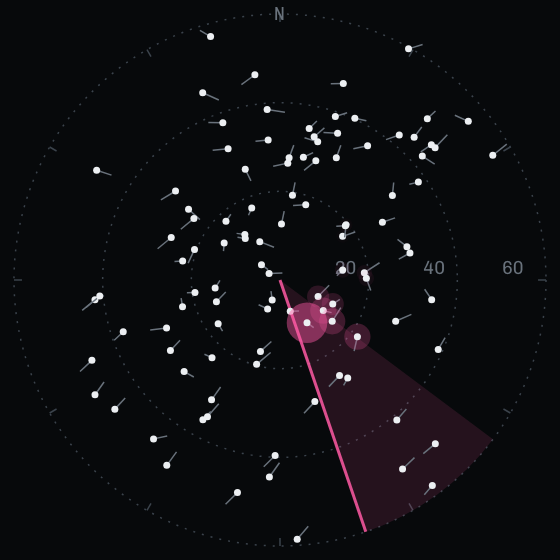}\\[2pt]{\small (a)}
\end{minipage}\hfill
\begin{minipage}[b]{0.43\linewidth}\centering
\includegraphics[width=\linewidth]{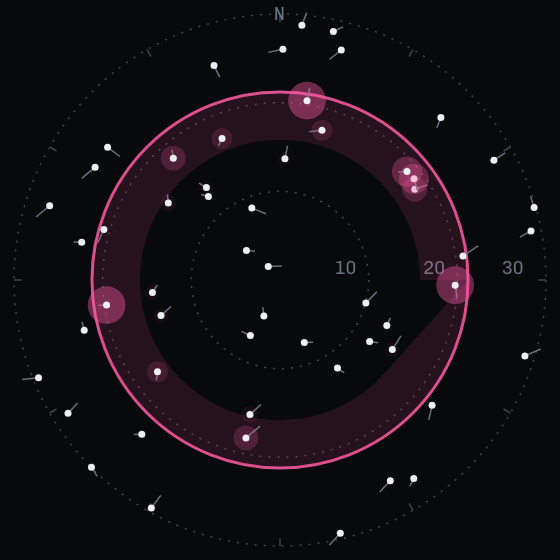}\\[2pt]{\small (b)}
\end{minipage}
\caption{The scope of the reference implementation while a scan runs, with the receiver at the centre and north at the top. (a) Scan by bearing, in the Radar director example: the line, and the fading wedge it has just swept; aircraft it has reached glow. (b) Scan by range, in the Range ping example at a 30 nautical mile scope: the ring, and the band it has just crossed.}
\label{fig:scopes}
\end{figure}

Take the defaults and a scan by bearing. The line starts due north and turns clockwise once every five seconds. An aircraft due east, at bearing 90 degrees, fires 1.25~s into each turn, since 90 is a quarter of 360 and a turn lasts five seconds. Now take a scan by range with a reach of 30 nautical miles and the same five-second turn. The ring starts at the receiver and grows to 30 nautical miles over the turn, so an aircraft at 15 nautical miles fires 2.5~s into each outward turn. The scope of the reference implementation draws the line or the ring while the patch plays (Figure~\ref{fig:scopes}).

That is the intended behaviour. What the reference implementation does is slightly different, because it does not watch for the crossing continuously. The patch is evaluated every 25~ms, and each evaluation looks 120~ms ahead on the audio clock so that notes can be scheduled at exact times. Between one evaluation and the next the beat clock advances by a small step, and the question asked at each evaluation is which aircraft the scan passed during that step. For a scan by bearing, the line turns through $360$ degrees times the step divided by the length of a turn, and every aircraft whose bearing lies inside that angle fires. For a scan by range, the ring moves through the same fraction of its reach, and every aircraft whose range lies inside that band, and not farther out than the reach, fires. An inward ring is handled by the same test with each aircraft's range fraction replaced by one minus that fraction, so the farthest aircraft fire first. In both cases the event is not stamped with the evaluation's time but placed inside the step by linear interpolation, in proportion to how far through the swept angle or band the aircraft lies.

Three limits are worth stating plainly. First, the crossing is tested once per 25~ms evaluation, using the aircraft positions at the data time of that evaluation, and not at the instant of crossing. Second, the scan front has no width, so there is no angular or range window on it: an aircraft has either been passed or not. Third, the Radar sweep is only one of several blocks that can start a note: Every, Pulse, Threshold, Rate limit, Enter/exit, Level-off, Closest pair and Ground also give events, and none of them is a scan. Changed starts no note of its own; it thins the events of another block.

\section{Musical choices}\label{sec:music}

The project began from music theory: the author wanted aircraft data to take musical form, with a pitch from altitude, a place in the bar from bearing and a loudness from signal strength. When two aircraft sound at once their pitches form an interval, and the scale decides which intervals can occur, since the Scale block snaps every value to a note of it. The editor offers nine scales: major and minor pentatonic, major, natural minor, dorian, whole tone, chromatic, and two derived from the minimum vectoring altitudes of Newark and of the Valle de M\'exico, the lowest altitudes at which a controller may steer aircraft over the terrain there. Counting the intervals between the notes of each, only the two pentatonic scales contain no minor second, major seventh or tritone. The whole-tone scale has a tritone and no minor second. The Newark scale, which is the default of the Scale block, has a minor second (E to F) and no tritone, and the other five have both.

The Flight levels example (Section~\ref{sec:levels}) uses the chromatic scale at one semitone per 1,000~ft on purpose, and the choice has a cost. (A flight level is an altitude quoted in hundreds of feet: FL180 is 18,000~ft.) In 94 percent of the frames of capture A and 99 percent of capture B, at least two of its ten tones stand on adjacent levels, and 27 and 25 percent of all pairs fall in the minor second, tritone or major seventh classes. At two semitones per level that share is 16 percent, and at three it is 25 or 26 percent. At four it is 0.2 and 0.3 percent. Two aircraft that stand on levels then form only a unison, a major third or a minor sixth, and the rest come from pairs in which an aircraft's altitude is far enough from a level to round to a different note. But only three pitch classes remain, and FL180 to FL450 would span nine octaves, so the selection would have to be narrowed. Only one semitone per level keeps the odd and even levels in the two whole-tone scales. A script kept with the source of this note, \texttt{check\_intervals.py}, reproduces these counts from the editor's scale table and the captures.

The Duet example (Section~\ref{sec:impl}) takes the idea of an earlier offline piece, in which two converging aircraft are two voices. Vertical separation sets the interval: on a minor pentatonic scale at 1,000~ft per step the legal vertical minimum, the least vertical distance required between two aircraft that are not separated in another way, is one step, and the scale has no tritone, so one can be sounded on purpose. The pace of the restrikes follows the time left to the pair's closest point, from 2.5~s between strikes at 90~s to go to 0.23~s at 5~s or less, with the example's settings at 96 beats per minute. If the pass is predicted inside 3 nautical miles and under 900~ft, the example sounds one tritone dyad, C and F sharp. The offline piece differed in two ways that matter: its pace followed the lateral distance, and it cut each encounter out with the closest point already known, whereas the block can only predict. We have not compared the two by listening.

The Ground block supplies a harmony where Scale supplies a scale: Scale snaps an altitude to a note of one fixed scale, Ground to a tone of a chord that changes as the piece goes on. The block holds a cycle of triads in a key, A minor by default, and offers four cycles, written as the degrees of the key on which each chord stands: i VII VI V, i V, i VII VI VII, and i VII i VI i V. Every chord is built on the natural minor except V, which is E, G sharp and B and so borrows from the harmonic minor its leading tone, the G sharp one semitone below the key note. The chords are a composed cycle: the data determine when the chord changes, which aircraft sound and on which chord tones, but not which chord comes next.

An aircraft in level flight sounds the chord tone nearest its altitude, in a window of three octaves spanning 0 to 12,000~ft by default. An aircraft climbing or descending faster than 300~ft per minute sounds instead a passing note, the scale note next to that chord tone, above when climbing and below when descending.

The nearest tone can move a level aircraft in step with the bass. Over the cycle i VII VI V, an aircraft at 5,750~ft sounds E, D and C a twelfth above the bass notes A, G and F, so it moves in parallel fifths with the bass at two of the four changes, a motion the rules of counterpoint forbid. A setting of the block, \emph{chord tones}, offers a second choice, \emph{voice-led}. At each change, the block chooses the tones of the aircraft in level flight that were sounding just before it as one combination. Each aircraft may take any chord tone within 3.5 semitones of its altitude, and the block keeps the combination with the least total motion, counting each parallel fifth or octave with another aircraft or with the bass as 6 semitones of motion. The aircraft at 5,750~ft then sounds E, B, C and E. The choice keeps two rules: a higher aircraft never sounds a lower note, and two aircraft within 300~ft of each other that the nearest tone gives one note keep one note. Between changes an aircraft keeps its tone while it stays level and within 3.5 semitones of its altitude, and an aircraft that is climbing or descending sounds its passing note in either setting.

The chord changes when an event reaches the block's step input, with two guards by default: no change in the first four bars after play starts, and none sooner than two bars after the last change. A Quantize set to bars in front of the step input lands the change on a bar line. A bass output gives the root of the chord near the bottom of the window, for a Voice struck once a bar, and a chord output gives the root, fifth and third as three notes, for a Voice that strums them.

Because every chord is a triad, no two aircraft in level flight sound a minor second, a tritone or a major seventh against each other within one chord. A passing note can: of the 16 cases of a chord and one of its passing notes, 10 make one of these intervals with a tone of the chord, so the statement is limited to level flight. It does not cover a note still ringing from the previous chord when the chord changes. \texttt{check\_intervals.py} reproduces both counts from the editor's chord table, and fails if a chord tone pair makes one of these intervals.

These are classifications from music theory and not listening results. No listener has been tested. The only listening judgements are the author's: that a minor second between two sustained tones sounds harsh in this setting, and that the Ground example (Section~\ref{sec:impl}) sounds somewhat musical. The second is the judgement of one person who built it, and not a listening test. In earlier offline pieces the author also judged a pure mapping of aircraft state to musical parameters informative but not musical. The blocks offer those mappings and the events, one relation between aircraft (Closest pair), and in Ground a harmonic frame inside which the data determine the notes and the moment of a change; beyond its four chord cycles they leave the form of a piece to the user.

\section{The reference implementation}\label{sec:impl}

\begin{figure}[t]
\centering
\includegraphics[width=\linewidth]{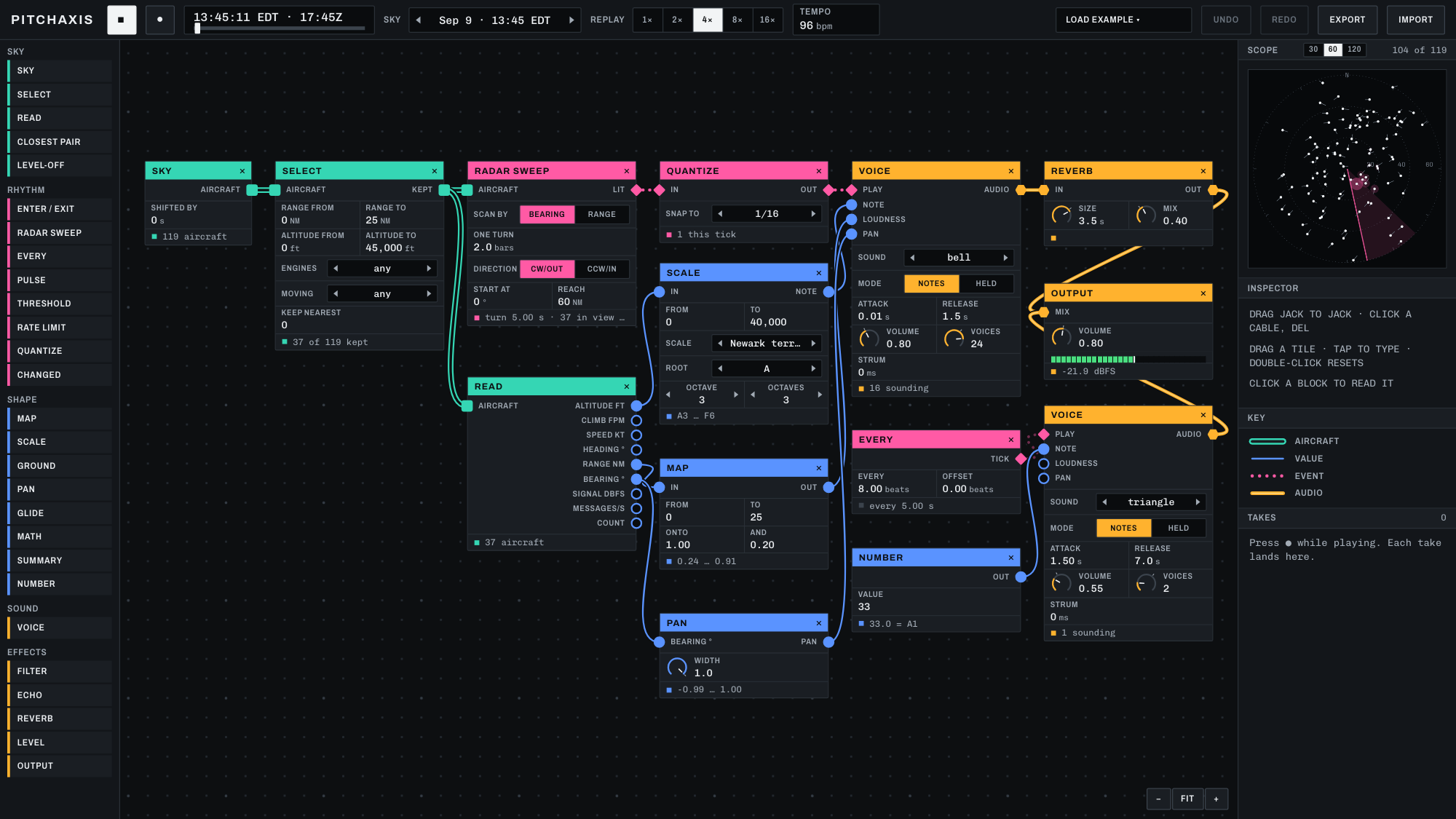}
\caption{The reference implementation with the Radar director example playing. Left, the library of blocks in five groups. Centre, the patch, with cables coloured by kind: aircraft set teal, value blue, event pink and dotted, audio amber. Right, the scope, the inspector, the key to the cables and the recorded takes.}
\label{fig:editor}
\end{figure}

The reference implementation is one HTML file of 1.3~MB. It contains the editor, the sound engine and the two captures, and it needs no server. Figure~\ref{fig:editor} shows it with one of the examples playing. Nothing in Sections~\ref{sec:model} and~\ref{sec:scan} depends on it. Another implementation would need a source of aircraft sets with the fields of Section~\ref{sec:signals} at a known time, a beat clock, and a sink that accepts notes with a time, a pitch, a loudness and a stereo position. The MIDI take described below is such a sink already. One further requirement comes from the model itself: blocks must be evaluable in an order in which each follows the blocks that feed it, which the refusal of loops guarantees. We have built only the browser implementation. Earlier offline versions of this work, written in Python, share some block names and ideas but are not instances of this model: one is a declarative registry with no graphical editor and a different set of kinds, and another lights seeded cells of the sky instead of crossing aircraft. The Duet example re-expresses a third, the offline conflict duet (Section~\ref{sec:music}).

\subsection{The data}

The reference implementation reads no live feed; it carries two recorded captures made on 9 September 2026 with one ADS-B receiver in central New Jersey, running readsb (Section~\ref{sec:signals}). Each capture is 355 frames, about 30 minutes. Capture~A starts at 17:45 UTC (13:45 EDT) and capture~B at 15:13 UTC (11:13 EDT). readsb applies plausibility checks before it writes a row~\cite{wiedehopf2026readsb}.

A packing step converts each capture into a compact replay. From each row it keeps the barometric altitude (altitude worked out from air pressure), the climb rate (the barometric one, or the GPS one when the barometric is absent), the ground speed and the track (the direction the aircraft is moving over the ground). It also keeps the range and bearing from the receiver (its distance and its compass direction), the signal strength, and a message rate worked out from the receiver's message count. For each aircraft it keeps the callsign and the aircraft type, and looks up the engine class and wake class from the type. Latitude and longitude are not kept.

The receiver goes on listing an aircraft after its last position message~\cite{wiedehopf2026readsb}, so a row can carry a position older than its frame. A row is kept only if its position is less than 6~s old and it has both a range and an altitude. Table~\ref{tab:captures} gives the counts. About a third of the rows are dropped, and the replay therefore holds fewer aircraft than the receiver listed. This is not the 73 percent of Section~\ref{sec:signals}: that figure comes from a different recording and counts an aircraft as heard if anything was received from it in the last 5~s, whereas the packing rule is about the age of the position and also needs a range and an altitude.

\begin{table}[h]
\centering\small
\begin{tabular}{@{}lrrrrrr@{}}
\toprule
Capture & Frames & Gaps of 5.0 s / 6.0 s & Rows seen & Rows kept & Aircraft seen & Aircraft kept\\
\midrule
A (17:45 UTC) & 355 & 330 / 24 & 54,525 & 36,688 & 407 & 396\\
B (15:13 UTC) & 355 & 328 / 26 & 48,531 & 32,509 & 379 & 368\\
\bottomrule
\end{tabular}
\caption{The two bundled captures and what the packing step keeps. Each has 354 gaps between frames.}
\label{tab:captures}
\end{table}

Frames are mostly five seconds apart, with some six-second gaps (Table~\ref{tab:captures}). Played back as recorded, every aircraft would jump once per frame, so the browser places each aircraft between its two nearest samples by linear interpolation, in proportion to the time elapsed, and turns bearing and track by the shortest way round the circle. If two samples are more than 20~s apart, the aircraft is held at the nearer sample for up to 2.5~s and then disappears. The default replay speed is four times real time, and the user can choose one, two, four, eight or sixteen.

\subsection{Working with the tool}

Every parameter on a block is shown as a tile, and every tile is handled the same way. The user drags it up or down to change the value, holds Shift to move in fine steps, taps it to type a value, and double-clicks it to restore the default. What the tile looks like follows from the shape of the parameter and not from the block it sits on. A parameter with at most four short options is a row of buttons, and one with more options is a picker. A number with both limits and at most seven steps between them is a stepper, any other number with both limits is a knob, and every other number is a plain tile. New blocks are placed in free space on the canvas. Edits can be undone and redone, each step restoring the whole patch. The editor keeps the last patch in the browser, so it is there again when the page is reopened.

The examples Radar director, Range ping, Approach chimes, Sky drone, Flight levels and Duet each load as a working patch, and so do seven built around Ground. Radar director scans by bearing. It takes pitch from altitude on a five-note scale, loudness from range and stereo position from bearing, plays the notes on a bell, and adds a long low triangle note on every eighth beat. Range ping scans by range over the aircraft within 30 nautical miles, so that the notes start sparse near the centre and become denser as the ring grows. Approach chimes uses Enter/exit and Threshold instead of a scan. Sky drone holds one continuous tone for each of the nearest aircraft. Duet plays the pair about to pass closest as the two voices that Section~\ref{sec:music} describes, and opens on a Boeing 737 MAX and a Cessna 206 in the first capture.

The Ground example has few voices over a composed harmony. A Select keeps the aircraft within 30 nautical miles and below 12,000~ft, and Level-off keeps the last three of these to settle level and, at the start, the three nearest that are already level: these are the cast. The cast sings the chord tones nearest its altitudes on a plucked voice, each note struck as a Radar sweep of two bars reaches its aircraft, and a bell sounds on the beat when an aircraft levels off. A bass is struck once a bar.

The chord changes at the next bar line after an aircraft descends through 4,000~ft within 30 nautical miles, and the cast then strikes its notes in the new chord; the example opens 280~s into capture~A, shortly before the first such descent. The aircraft that triggers the change is read by a Select of its own, so that it counts whether or not it is one of the voices.

The other six Ground examples are variants made to listen against it, each changing one thing. In \emph{on a clock}, the chord changes every 24 beats instead of when an aircraft descends. In \emph{shifted sky}, it changes on the descents of the same sky 300~s later, so that the timing has the same kind of source but is not these aircraft's. In \emph{strummed}, the bass bar is the whole chord, rolled, instead of the root alone. In \emph{thinned}, a Changed block stops an aircraft's note being struck again unless that note has changed or four bars have passed. In \emph{voice-led}, Ground chooses its chord tones voice-led instead of nearest (Section~\ref{sec:music}). In \emph{cast only}, there is no Ground, and the same cast and bells play on the Newark terrain scale with the long low note of the Radar director. This note reports no comparison among the seven. Flight levels is described next.

\subsection{Sound, recording and the MIDI take}

A Voice block makes the sound. It has seven timbres: bell, pluck, sine, triangle, sawtooth, square and a band of noise. It works in notes mode or in held mode. In notes mode it plays a short note for each event it receives, with an attack and a release time. In held mode every aircraft in the set holds its own continuous tone, and the tone slides to a new pitch when its note value changes. A polyphony limit caps how many tones can sound at once. Filter, Echo, Reverb and Level shape the sound on its way out, and the Output block applies a gentle limiter, which holds the loudest peaks below a ceiling, shows a level meter, and sends the sound to the speakers.

While the patch plays, a Record button captures what leaves the Output block. When the take ends, the page offers two files. The audio file is the sound itself, as WebM or MP4 depending on the browser. The MIDI file holds no sound; it is a list of the notes the Voices played in the same window, each with its pitch and its start and end times, so the notes can be given any other instrument in a music program. It is a Standard MIDI File of format 1 at 480 ticks per quarter note. Each Voice has its own track on its own channel, and channel~10, reserved for percussion, is never used. Each track starts with a General MIDI program chosen from the Voice's timbre: tubular bells for bell, nylon guitar for pluck, a pad for sine, flute for triangle, a sawtooth lead for sawtooth, a square lead for square and sea shore for noise. Stereo position is written as controller~10, and the tempo is taken from the clock when recording starts. We read the files back with an independent MIDI library. They parsed without error and every note that started also ended. For the two examples that strike notes, the file held the same number of notes as the Voices played in the recording window.

\section{A worked mapping: flight levels}\label{sec:levels}

\begin{figure}[t]
\centering
\includegraphics[width=\linewidth]{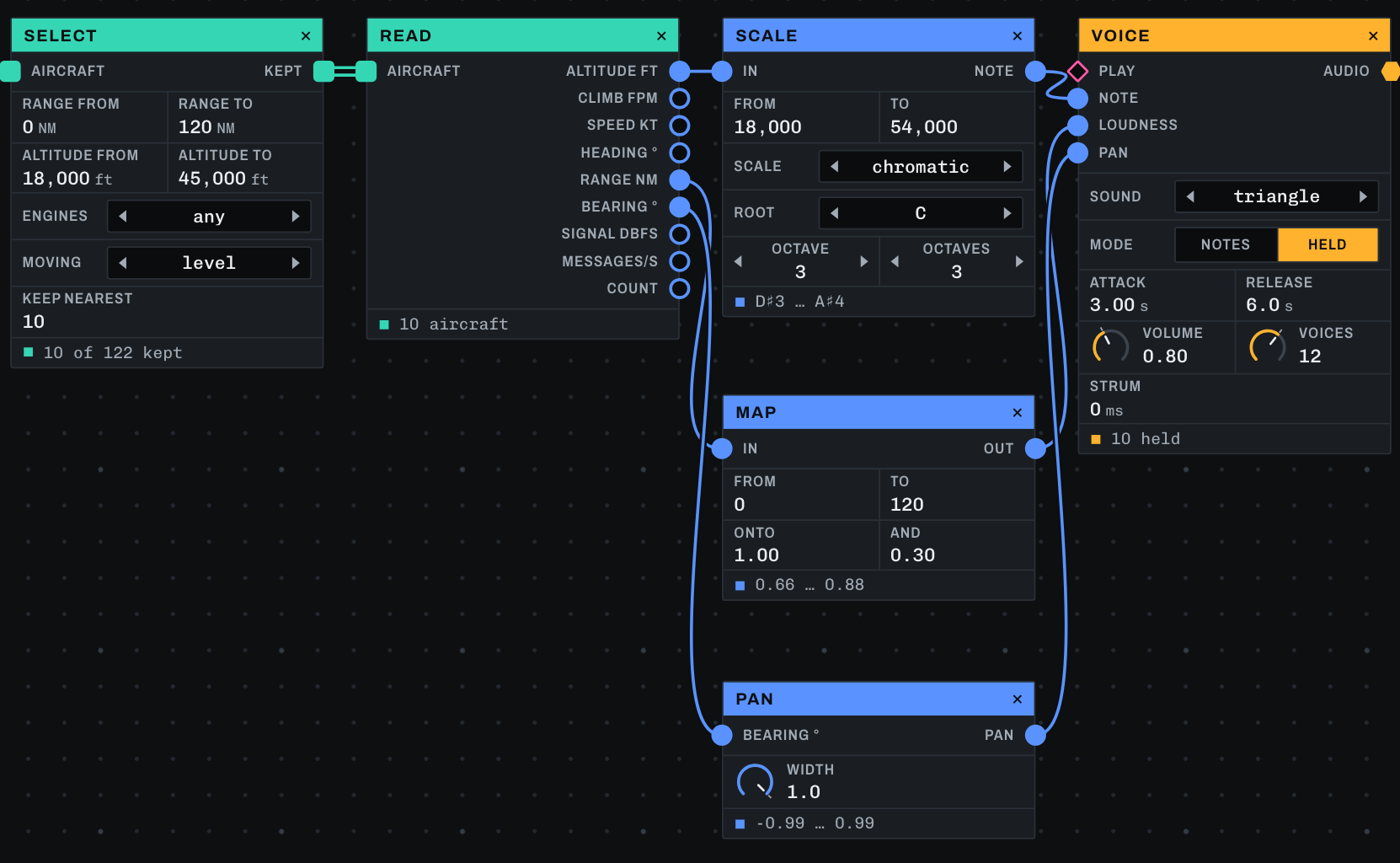}
\caption{The Flight levels example playing, from Select to Voice. Select keeps the ten nearest aircraft in level flight between 18,000 and 45,000~ft within 120 nautical miles. Read opens altitude, range and bearing. Scale sets a chromatic scale from 18,000 to 54,000~ft; Map turns range into loudness and Pan turns bearing into stereo position. A held Voice gives each aircraft a triangle tone.}
\label{fig:levels}
\end{figure}

Cruising aircraft fly at flight levels. A flight level is an altitude named in hundreds of feet, so FL180 is 18,000~ft, and it is measured against one standard pressure setting rather than the local one. The Flight levels example keeps the ten nearest aircraft that are in level flight between 18,000~ft and 45,000~ft, where level flight means a climb rate of at most 300~ft per minute in magnitude, and holds each as a continuous triangle tone (Figure~\ref{fig:levels}). An aircraft that reports no climb rate is never counted as level, and 24 and 21 percent of the replayed rows in the two captures report none. Loudness falls with range and stereo position follows bearing.

The census gives the resolution that the mapping depends on. Altitude is stepped, with a grid of 25~ft (Table~\ref{tab:signals}), and flight levels are 1,000~ft apart. The pitch comes from altitude through a chromatic scale running from 18,000~ft to 54,000~ft over three octaves and starting at the note C3. Because 18,000~ft to 54,000~ft is 36,000~ft and three octaves hold 36 semitones, each 1,000~ft is one semitone, and each flight level therefore has its own note. The scale runs to 54,000~ft, above the 45,000~ft limit of the selection, so that no aircraft is clamped to the top note.

We chose this mapping on purpose, and it shapes how the result sounds. Adjacent flight levels alternate between odd and even thousands of feet, and since the scale starts at 18,000~ft, odd levels fall on odd semitones and even levels on even semitones; each group is one of the two whole-tone scales. The thirteen levels from FL290 to FL410 span exactly one octave. And two aircraft on adjacent flight levels are a minor second apart. Our earlier design notes call one semitone per 1,000~ft a trap for exactly this reason: aircraft with legal vertical separation then sound a minor second apart. That is why the default Scale mapping is different, and that default is unchanged.

The mapping rests on one premise: that cruising aircraft really sit on flight levels, so that one semitone per 1,000~ft gives one note per level rather than a smear of pitches in between. We checked the premise on the captures. A level row is a row with a barometric altitude of at least 18,000~ft and a climb rate of at most 100~ft per minute in magnitude, taken from the raw rows before the packing step; we count rows, not aircraft. Table~\ref{tab:levels} gives the result. Of the level rows, 99.8 and 99.6 percent lie within 100~ft of a multiple of 1,000~ft. If altitudes were spread evenly, 22.5 percent would: the altitudes come on a 25~ft grid, and of the forty grid points in each 1,000~ft, the nine that run from 100~ft below a level to 100~ft above it pass the test.

\begin{table}[h]
\centering\small
\begin{tabular}{@{}lrrrr@{}}
\toprule
Capture & Level rows & Within 100 ft of a level & Share & Aircraft behind the rows\\
\midrule
A & 8,447 & 8,433 & 99.8\% & 128\\
B & 7,892 & 7,861 & 99.6\% & 110\\
\bottomrule
\end{tabular}
\caption{Level rows at or above 18,000 ft and how many lie within 100 ft of a flight level. Chance is 22.5\%.}
\label{tab:levels}
\end{table}

We present this as a check, not a finding, for three reasons. First, the clustering is what the rules already require. Aircraft in level flight are meant to be on their assigned levels; above 18,000~ft pilots set the altimeter to 29.92 inches of mercury, the same standard setting for everyone~\cite{cfr2026s91121}; and the altitude-keeping requirement in airspace with reduced vertical separation is $\pm$65~ft, or $\pm$130~ft for some older aircraft~\cite{cfr2026appg}. In one study of recorded flights, cruise altitudes stayed within about 12~m of the level~\cite{chen2014relative}. Second, every altitude in the captures is a multiple of 25~ft, which is how Mode S, the transponder format in which the altitude is sent, encodes it~\cite{sun2021riddle}; this says something about the decoder and nothing about the aircraft. Third, ADS-B climb rate comes in steps of 64~ft per minute~\cite{wiedehopf2026readsb}, so our test at 100~ft per minute admits only rows reporting zero or plus or minus 64. There are further cautions. The check uses a stricter climb-rate limit than the example does (100~ft per minute against 300); at 300~ft per minute the shares are 99.5 and 98.7 percent, and on the replayed rows rather than the raw rows at 100~ft per minute they are 99.95 and 99.81 percent. A script kept with the source of this note, \texttt{check\_flight\_levels.py}, reproduces Table~\ref{tab:levels} and these figures from the captures. The rows are not independent, since each aircraft contributes many of them, and the captures come from one receiver on one day. We did not measure the semicircular rule, which puts eastbound traffic on odd flight levels~\cite{cfr2026s91179}. Jansen plotted the rule qualitatively on Flightradar24 data~\cite{jansen2020flightradar24}.

Mapping altitude to pitch is common, and a published guideline rounds a linear map of a data range to the nearest MIDI note~\cite{brown2003guidelines,dubus2013mapping}. The part we add is the choice of constant and anchor, 1,000~ft per semitone starting at FL180, which lines the notes up with the flight levels.

\section{Relationship to earlier work}\label{sec:related}

We looked for earlier systems in October 2026 among papers, code repositories and project pages that we could read. The search did not cover video platforms, forums or patents, so every statement below that something was not found is limited to the sources we reached. Table~\ref{tab:related} lists the closest systems, what each does, and what we did not find in it. Four terms in it may be new: ASTERIX is a standard format for radar data, an earcon is a short abstract sound that carries a message, adsb.lol is a public service that provides ADS-B data, and Voltage Modular is a software modular synthesizer whose modules are joined by cables.

\begin{table}[p]
\centering\footnotesize
\newcolumntype{L}[1]{>{\raggedright\arraybackslash}p{#1}}
\begin{tabular}{@{}L{4.0cm}L{0.9cm}L{5.2cm}L{4.2cm}@{}}
\toprule
System & Year & What it does & What we did not find in it\\
\midrule
SYNTHRADAR~\cite{ehjay2026synthradar} & 2026 & A radar scope over adsb.lol aircraft with an expanding range ring, and a synthesizer patch for each aircraft. On listening, a note begins as the ring reaches an aircraft. & A user-facing editor or configurable modes. A bearing scan. Public code.\\
scoreTable*~\cite{jorda2006scoretable} & 2006 & A radar sweep over camera-tracked pucks that triggers a note when it passes a puck, with global pucks for the tempo and for the number of bars in one tour of the sweep, from one to eight. & Aircraft or any other moving data. A range ring.\\
ProjectRadarSequencer~\cite{hellicar2013radarsequencer} & 2013 & Positions (from laser scanners in its description, from mouse drags in its demo) are mapped to angular columns, with one revolution per measure derived from the tempo. & Aircraft. An editor.\\
RhythmTable~\cite{williams2026rhythmtable} & 2026 & A tangible rhythm sequencer whose rotating playhead triggers audio as it crosses each token's angle. & Aircraft or data.\\
Cassidy et al.~\cite{cassidy2019radar} & 2019 & A virtual radar arm that gives directional sound cues, with sweeps in 3/4 and 4/4 time, tested with 36 participants. & Tracked moving objects.\\
SonAir~\cite{elmquist2023sonair} & 2023 & Continuous directional sound for each aircraft plus an entry earcon, for air traffic controllers, using simulator ASTERIX data every fifth second. & A scan: onset follows each data arrival. ADS-B data. An editor.\\
Home Assistant radar card~\cite{ehrenholm2026radarcard} & 2026 & A rotating sweep over ADS-B aircraft with a bearing-crossing test like ours. & Sound from the sweep: the crossing brightens a blip, and sound alerts follow other events.\\
adsbTheremin~\cite{good2015adsbtheremin} & 2015 & Replays recorded dump1090 logs at adjustable speed (since 2022) and writes a MIDI file (since August 2026). & A scan front. An editor.\\
Flight-DataSonification~\cite{raddad2026flightdatasonification} & 2026 & A browser sonification of flight data that records audio and raw packets. & A scan front. An editor.\\
Sonar Sequencer~\cite{playertron2026sonarsequencer} & n.d. & A module for Voltage Modular: a green line sweeps a circular path past markers on concentric rings and outputs trigger pulses. & Aircraft or other tracked data (its store page mentions none).\\
PingSeq~\cite{mooy2014pingseq} & 2014 & A sequencer for Max, a visual patching environment: each time the sweep's angle meets a node's angle it sends a list of that node's values, and the user's patch decides the sound. & Aircraft or other tracked data (its README mentions none).\\
NASA Chandra X-ray Center~\cite{nasa2020sonification} & 2020 & A sonification of an image of the Galactic Center: a sweep from left to right, with vertical position setting pitch and brightness setting volume; stars sound as notes and gas as a drone. & Aircraft or tracked moving objects. A rotating or ring-shaped front. An editor.\\
libmapper~\cite{libmapper2026} & 2026 & Typed connections between signals, made in its graphical editors; a signal can carry a changing set of tracked things. & Aircraft. A scan block.\\
\bottomrule
\end{tabular}
\caption{The closest systems we found and what each does. The right-hand column is limited to what we could read.}
\label{tab:related}
\end{table}

Both halves of the scan were published earlier by others. The rotating line, a radar sweep locked to a tempo and passing over tracked objects, is old: scoreTable* published it for camera-tracked pucks in 2006, and ProjectRadarSequencer did the same for tracked positions in 2013. The ring, a range ring over ADS-B aircraft that starts a note as it reaches each aircraft, was published first by SYNTHRADAR, on a platform date of 30 September 2026. We implemented our range scan on 16 October 2026, after we had seen it. We do not claim to be first to sweep a ring or a line across aircraft with sound.

A sweep that triggers sound has also been placed inside patching environments: Sonar Sequencer in a modular synthesizer and PingSeq in Max. A left-to-right sweep over an image of the sky, with position setting pitch and brightness setting volume, was used by the Chandra X-ray Center for the Galactic Center in 2020. Scanning is not new in any of these settings.

What we did not find elsewhere is narrower than a scan block in a typed editor: the aircraft set as a kind of cable. One connection carries every aircraft in view with all of its fields attached. Blocks open it by field (Read) or filter it by aircraft (Select), and the numbers derived from it are keyed by aircraft, so that they meet again at one voice per aircraft. A scan inside a patching environment, and typed connections, are not new, as the systems above, libmapper~\cite{libmapper2026} and Max, a visual programming environment for music~\cite{cycling74msp}, show, so we do not claim the combination of a scan block and a typed editor. The polyphonic cables of modular environments also carry several channels in one connection, and we have not compared them with the aircraft set. We make no claim that the aircraft set is better for any task. Controllers in one operational alert scheme reported desensitization and annoyance when high-urgency alarms were frequent~\cite{cabrera2006considerations}, and we do not claim that a scan reduces workload.

One further project is listed but unread. A doctoral project at the German Aerospace Center on sonifying air traffic control data, titled ``Underscoring Aircraft Ballets: Sonifying Air Traffic Control Data to heighten Situational Awareness: A Composer's Perspective'', appears in a 2021 biography~\cite{vondernahmer2021bio}. We found no thesis, paper or description of its method, so we cannot compare it with our work.

\section{What is not built}\label{sec:limits}

Several things that we have discussed are not built.
\begin{itemize}
\item \textbf{Scan shapes and control.} The scan front has no angular or range window, and there is no spiral or back-and-forth scan. There is no way to steer the scan by hand, with a mouse or a finger.
\item \textbf{Types beyond the kind.} Value cables carry no unit or range, so an altitude or a speed can be set below zero and the editor does not stop it, and a cable does not say that its numbers are angles; Section~\ref{sec:model} describes the two blocks that then treat a bearing wrongly. The design we intend is that a type supplies limits, units and warnings and never rejects a cable or rewrites a value the user has set.
\item \textbf{Signals the tool does not carry.} The census describes more than the tool carries: the values the pilot has selected, the transponder code, the source of the contact and the quality codes are not in the tool's data.
\item \textbf{Relations between aircraft.} There is one block for a relation between two aircraft, Closest pair, and it gives only the single pair about to pass closest; every other block acts on one aircraft at a time or on the whole set. It predicts in a straight line from ground speed and track, so it is wrong in turns and level-offs, and positions are up to 6~s old. A predicted close pass is a prediction from recorded positions, not a statement about separation: visual separation and parallel approaches are legal inside 3 nautical miles and 900~ft. Nobody but the author has judged the Duet example by listening.
\item \textbf{Harmony.} Ground has four chord cycles, all of triads, in one key at a time. The data determine when the chord changes and which aircraft sound, not what chord comes next, so the form of a piece is the user's. The voice-led setting avoids parallel fifths and octaves only among the chord tones within 3.5 semitones of each altitude, so a parallel can remain, for example where only one tone is in reach. Its cost counts only motion and those parallels, so nothing stops it choosing a leap of a tritone. Level-off announces that an aircraft has levelled off only after it has stayed level for a set time, 10~s of data time by default, because it waits to be sure; a note struck from that event therefore sounds after the moment it marks.
\item \textbf{Two time bases.} Glide counts seconds of real time, and Enter/exit and Level-off count seconds of data time, so they disagree at any replay speed other than one.
\item \textbf{Data.} The editor cannot load data from the user's own receiver or from a live feed; it carries only the two bundled captures. The census statistics come from one recording at one receiver.
\item \textbf{Repeatability.} Rate limit starts each aircraft's timer at a random phase, so when its first events fire differs from one play to the next. The noise timbre and the reverb also use random numbers. No other block's timing is random.
\item \textbf{Evidence.} The tool has been used and tested only by its author, and no user has been observed using it. The only listening judgement of the Ground example is the author's. The Closest pair, Pulse, Level-off, Changed and Ground blocks and the Duet and Ground examples have a committed test script, \texttt{test-duet.cjs}, that runs in Chromium without a window (headless); the scan block has no committed test. The figures in Table~\ref{tab:levels} are reproduced by a script, \texttt{check\_flight\_levels.py}, that is kept with the source of this note and is not part of the tool.
\end{itemize}
The statements in this note about work we did not find are limited to the sources we reached and are worded ``to our knowledge'' for that reason.

\section{Availability and use of AI}

The code, the built page, the two captures, the tests and the source of this note are at \url{https://github.com/arquicanedo/patchbay}. The code is under the MIT License; this note, its figures and the captures are under CC BY 4.0.

This work was done with Claude Code and Claude models (Sonnet~5.5, Opus~5.5 and Fable~5.1), which wrote the code and drafted this note under the author's direction. Measured from the session logs of 14 to 25 October 2026 with the author's effort-meter tool, the editor and this note took 137 prompts from the author (2,422 words) and about 2.3 hours of the author's interaction, counting at most two minutes per reply, against 15.6 hours of agent work and 2.5 million output tokens. Time away from the keyboard and the earlier work on the captures and the census are not counted.

\bibliographystyle{plainurl}
\bibliography{references}

\end{document}

%% file: figures/signals-table.tex
\begin{tabular}{@{}llp{7.1cm}r@{}}
\toprule
Signal & Shape & Range and resolution, or number of values & Present (\%)\\
\midrule
\multicolumn{4}{@{}l}{\emph{Made by the receiver}}\\
Signal strength & continuous & $-$24.9 to $-$2.7 dBFS & 100\\
Message rate & continuous & 0.2 to 21.0 msg/s & 89\\
Range & continuous & 3.7 to 90.1 NM & 86\\
Bearing & angle & 0 to 360 degrees, wraps & 86\\
\midrule
\multicolumn{4}{@{}l}{\emph{Broadcast by the aircraft}}\\
Altitude (pressure) & stepped & 925 to 43,000 ft; steps of 25 ft & 98\\
Climb rate & stepped & $-$2,880 to 3,264 ft/min; steps of 64 ft/min & 74\\
Ground speed & continuous & 81 to 488 kt; whole-knot components & 98\\
Track & angle & 0 to 360 degrees, wraps & 98\\
\midrule
\multicolumn{4}{@{}l}{\emph{Set by the pilot or assigned by control}}\\
Selected altitude & stepped & 1,088 to 43,008 ft; steps of 32 ft & 90\\
Selected heading & angle & 0 to 360 degrees, steps of 0.70 degree & 60\\
Pressure setting & stepped & 1,012.8 to 1,016.8 hPa; steps of 0.8 hPa & 89\\
Autopilot modes & combination & six switches, 21 combinations seen & 23\\
Transponder code & category & 271 codes among 287 aircraft, 7.9 bits & 81\\
\midrule
\multicolumn{4}{@{}l}{\emph{Identity, unchanged while in view}}\\
Size class & category & 6 values, 1.7 bits & 94\\
Aircraft type & category & 77 values, 5.4 bits & 98\\
Operator & category & 95 values, 5.0 bits & 90\\
\midrule
\multicolumn{4}{@{}l}{\emph{How the contact arrived}}\\
Source of the contact & category & ADS-B 90\%, Mode S only 7\%, multilaterated 3\% & 100\\
Quality codes & combination & 42 combinations of eight codes, 2.4 bits & 88\\
\midrule
\multicolumn{4}{@{}l}{\emph{The whole sky}}\\
Frame clock & stepped & a grid of 5 s; longest gap 6.0 s & n/a\\
\bottomrule
\end{tabular}

%% file: figures/fig-kinds.tex
\begin{tikzpicture}[
  >=Stealth,
  kind/.style={rounded corners=2pt, draw, thick, minimum width=3.2cm, minimum height=1.3cm, align=center, inner sep=4pt},
  lab/.style={font=\scriptsize\sffamily, align=center, fill=white, inner sep=1.5pt},
  src/.style={font=\scriptsize\sffamily, text=black!70},
  arr/.style={->, thick, black!65}]
\definecolor{kA}{RGB}{20,150,128}
\definecolor{kV}{RGB}{60,110,230}
\definecolor{kE}{RGB}{225,50,120}
\definecolor{kU}{RGB}{225,150,20}
\node[kind, draw=kA, fill=kA!13] (A) at (0,0) {\textbf{Aircraft set}\\[2pt]\scriptsize\sffamily keeps its kind: Select,\\\scriptsize\sffamily Closest pair, Level-off};
\node[kind, draw=kV, fill=kV!13] (V) at (6.1,1.9) {\textbf{Value}\\[2pt]\scriptsize\sffamily keeps its kind: Map, Scale, Ground,\\\scriptsize\sffamily Pan, Glide, Math, Summary};
\node[kind, draw=kE, fill=kE!13] (E) at (6.1,-1.9) {\textbf{Event}\\[2pt]\scriptsize\sffamily keeps its kind: Quantize, Changed};
\node[kind, draw=kU, fill=kU!13] (U) at (13.4,0) {\textbf{Audio}\\[2pt]\scriptsize\sffamily keeps its kind: Filter, Echo,\\\scriptsize\sffamily Reverb, Level};
\draw[arr] (A.east) -- node[lab, pos=.5] {Read,\\Closest pair} (V.west);
\draw[arr] (A.east) -- node[lab, pos=.5] {Radar sweep,\\Enter/exit,\\Closest pair,\\Level-off} (E.west);
\draw[arr] (V.south) -- node[lab, pos=.5] {Threshold,\\Rate limit,\\Pulse} (E.north);
\draw[arr] ([xshift=0.35cm]E.north west) -- node[lab, pos=.5] {Ground:\\next chord} ([xshift=0.35cm]V.south west);
\draw[arr] (V.east) -- node[lab, pos=.5] {Voice:\\note, loudness, pan} (U.west);
\draw[arr] (E.east) -- node[lab, pos=.5] {Voice:\\play} (U.west);
\node[src] (sky) at (0,2.1) {Sky};
\node[src] (nb) at (6.1,3.85) {Number};
\node[src] (ev) at (6.1,-3.85) {Every};
\node[src] (out) at (13.4,-2.1) {Output: speakers and files};
\draw[arr] (sky) -- (A.north);
\draw[arr] (nb) -- (V.north);
\draw[arr] (ev) -- (E.south);
\draw[arr] (U.south) -- (out);
\end{tikzpicture}